\documentclass[a4paper]{jpconf}
\usepackage{graphicx}
\usepackage{sidecap}  
\usepackage{amsmath}
\begin{document}
\title{Quasi-particle current in planar Majorana nanowires}

\author{Javier Osca and Lloren\c{c} Serra}
\address{Institut de F\'{\i}sica Interdisciplin\`aria i de Sistemes Complexos IFISC (CSIC-UIB), E-07122 Palma de Mallorca, Spain}
\address{Departament de F\'{\i}sica, Universitat de les Illes Balears, E-07122 Palma de Mallorca, Spain}
\ead{javier@ifisc.uib-csic.es}

\begin{abstract}
We calculate the local quasi-particle current of a Majorana state in a planar hybrid (superconductor-semiconductor) nanowire. 
In absence of perpendicular components of the magnetic field the current flows in circular trajectories without a preferred orientation.
On the other hand, when a perpendicular component of the magnetic field is present the quasi-particle current circulates surrounding the Majorana density peak with an orientation established by the magnetic field. 
\end{abstract}

\section{Introduction}
Majorana modes appear in quasi-1D wires as effectively charge-less, zero-energy eigensolutions. They arise  from the splitting, through a phase transition, of bulk electronic states into pairs of quasi-particles on the wire ends, each one  being its own antiparticle \cite{Alicea,Beenaker2,StanescuREV,Franz}. These phase transitions are known to occur at particular
values of the magnetic field  \cite{Lutchyn2} giving rise to characteristic phase diagrams. Recently, we have investigated the role of the magnetic 
orbital motion
on the physics of Majorana states. In Ref. \cite{Osca4} it is shown that in a planar nanowire the main effect of the orbital motion is to change the Majorana phase boundaries. In general, stationary Majorana states can sustain 
non-vanishing local quasi-particle currents. 
Furthermore, these currents are altered by the kinetic orbital motion caused by the off plane components of the magnetic field. In this proceedings article we extend the work of Ref. \cite{Osca4} studying the quasi-particle current present in planar Majorana nanowires with and without components of the magnetic field perpendicular to the nanowire surface.

\section{Physical model}
\begin{SCfigure} 
\centering
\resizebox{0.45\textwidth}{!}{
	\includegraphics{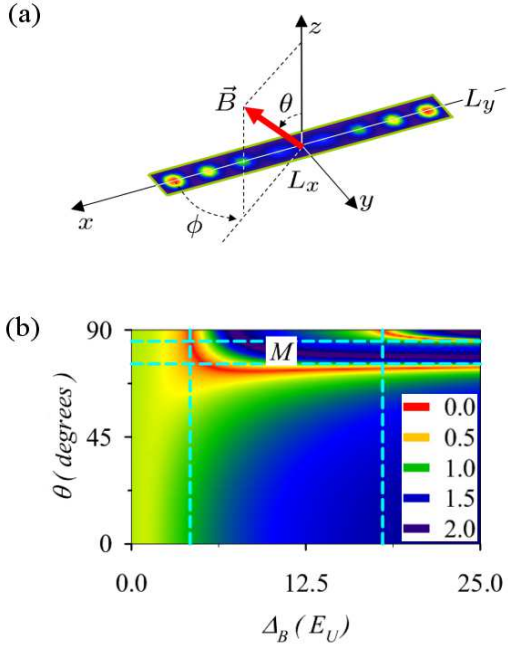}
}
\caption{ a) Schematic of a 2D planar nanowire showing the axis definitions. The density distribution of Majorana modes on the wire ends is qualitatively shown. b) Phase transition proximity measure with phase transitions (at zero value) in red, as a function of  $\Delta_B$ and polar angle $\theta$. The azimuthal angle remains $\phi=0$.  The results correspond to an InAs nanowire with $\alpha= 30\, {\rm meV}{\rm nm}$, $\Delta_s=0.3\, {\rm  meV}$, gyro-magnetic factor $g=15$, effective mass ratio $m^*=0.033$, and $L_y=150\, {\rm nm}$  in a magnetic field range from zero to 6 T. Both figures taken from Ref.\ \cite{Osca4}.}
\label{F1}
\end{SCfigure}

Majoranas can be obtained in nanowires due the combined effects of s-wave superconductivity, Rashba interaction and an external magnetic field. We consider a nanowire where the electronic motion is restricted to the $\hat{x}$ (longitudinal) and $\hat{y}$ (transverse) directions
in presence of these three effects. The homogeneous magnetic field points in an arbitrary direction and the edges are modeled as infinite square well potentials in the longitudinal and transverse directions (see Fig.\ \ref{F1}a). Therefore the nanowire physics is described by a Hamiltonian of the Bogoliubov-de Gennes kind
\begin{equation}
\begin{split}
\mathcal{H}_{\it BdG} &= \left( \frac{p_{x}^{2}+p_y^2}{2m} + V(x,y) -\mu \right)\tau_z + \Delta_s\, \tau_{x} + \frac{\alpha}{\hbar}\, \left(\, p_x \sigma_{y} - p_y \sigma_{x}\, \right)\tau_z \\
&+ \Delta_B\, \left(\sin\theta \cos\phi\, \sigma_x	+ \sin\theta \sin\phi\, \sigma_y +\cos\theta\, \sigma_z \right )\; , 
\end{split}
\label{E1}
\end{equation}
where the different terms are, in left to right order: kinetic energy, electrical potential $V$, chemical potential $\mu$, the superconductor term with strength $\Delta_s$,  the Rashba spin orbit interaction term with coupling strength $\alpha$ and finally, the Zeeman term of a magnetic field of magnitude $\Delta_B$ and arbitrary polar and azimuthal angles $(\theta,\phi)\equiv\hat{n}$. The Pauli operators for isospin (particle-hole) are represented by $\tau_{x,y,z}$ while those for spin are $\sigma_{x,y,z}$. In a planar nanowire the perpendicular component of the magnetic field induces orbital motions of the nanowire quasi-particles through the substitution  $\vec{p} \to \vec{p} + \frac{e}{c}\vec{A}(x,y)$ in the kinetic and Rashba terms (using the $e>0$ convention). The additional orbital terms
obtained with this substitution in Eq.\ (\ref{E1}) are given in Ref.\ \cite{Osca4}.

The semi-infinite nanowire exact zero-energy solution can be obtained by means of the complex band structure method of Refs.\ \cite{Osca4,Serra}. 
More specifically, the infinite nanowire eigensolutions are obtained numerically from the effective one dimensional Hamiltonian. Imposing the boundary condition to a superposition of infinite-nanowire solutions with different wave numbers yields the Majorana edge state. 
This method has the advantage of being computationally effective in comparison with direct diagonalization methods, even with fine spatial grids,  and can also be used to obtain the phase diagram of a Majorana nanowire inside a tilted magnetic  field as shown in Fig.\ \ref{F1}b. In particular, a phase transition proximity measure is displayed where phase boundaries (zero values) appear as red lines, and the regions with a Majorana mode are labeled with an M while regions without a Majorana are not labeled. Dashed lines indicate the phase transition analytical limits
\begin{equation}
 \Delta_{B,n}^{(c)}=\sqrt{\left(\mu -\epsilon_{n}+\frac{m\alpha^2}{2\hbar^2} \right)^2+\Delta^2_s}\;,
 \label{E28}
\end{equation} 
and 
\begin{equation}
 \theta^{(c)}_n=\arccos \left( \frac{gm^*}{4(n-\frac{1}{2})} \right)\;,
 \label{E29}
\end{equation}
where $n=1,2,3,...$ . We refer to Ref.\ \cite{Osca4} for further details. The results are presented in characteristic units of the problem obtained by taking  $\hbar$, $m$ and the width of the nanowire $L_y$  as reference values. That is, our length and energy units are, respectively, 
$L_{\it U}\equiv L_{y}$ and $E_{\it U}\equiv {\hbar^2}/{m L_{y}^2}$.

\section{Quasiparticle current}
The quasi-particle continuity equation for the Hamiltonian of  Eq.\ (\ref{E1}),
\begin{equation}
\frac{\partial \rho(x,y)}{ \partial t}=-\nabla\cdot\vec{\jmath\,}(x,y) \;,
\label{E30}
\end{equation}
is obtained from the Majorana probability density  $\rho(x,y)=\Psi(x,y)\Psi^*(x,y)$ and the time dependent Schr\"odinger equation $i \hbar \partial_t\Psi(x,y)=\mathcal{H}_{BdG}\Psi(x,y)$, where $\Psi(x,y)$ is defined as the four-component spinor $\Psi\equiv\left(\psi_{+ +},\psi_{- +},\psi_{+ -},\psi_{- -} \right)^T$ in spin and isospin space. The rate of change of the Majorana density depends on the divergence of the quasi-particle current  
\begin{equation}
\vec{\jmath\,}(x,y)=\Re\left[\, \Psi^*(x,y)\, \hat{v}\, \Psi(x,y)\, \right]\;,
\label{E31}
\end{equation}
that is directly proportional to the real part of the velocity operator 
$\hat{v}\equiv(\hat{v}_x,\hat{v}_y)$, where
$\hat{v}_x=\partial \mathcal{H}/\partial p_x$ and  $\hat{v}_y=\partial \mathcal{H}/\partial p_y$,
in the same way as in Ref. \cite{Molenkamp}. For our particular Hamiltonian the $x$ and $y$ components
of the velocity operator read
\begin{eqnarray}
\hat{v}_x &=& -i \frac{\hbar}{m} \partial_x \tau_z + \frac{e}{m c}A_x(x,y) + \frac{\alpha}{\hbar} \sigma_y \tau_z\;,
\label{E32}\\
\hat{v}_y &=& -i \frac{\hbar}{m} \partial_y \tau_z + \frac{e}{m c}A_y(x,y) - \frac{\alpha}{\hbar} \sigma_x \tau_z\;.
\label{E34}
\end{eqnarray}

As the density is constant in time, the resulting current has zero divergence,
 $\nabla\cdot\vec{\jmath}=0$.
Equation (\ref{E31}) can be rewritten as the sum of a familiar quasi-particle current for non Rashba superconducting devices \cite{Vignale,Beenaker}, plus a Rashba current $\vec{\jmath}_{so}$
\begin{equation}
\vec{\jmath\,}(x,y)=\frac{\hbar}{m}\,
\Im\left[\, \Psi^*(x,y)\, \nabla \tau_z\, \Psi(x,y)\, \right]
+\frac{e}{m c}\,\rho(x,y)\,\vec{A}(x,y)+\vec{\jmath}_{so}(x,y)\;,
\end{equation}
where
\begin{equation}
\vec{\jmath}_{so}(x,y)=\frac{\alpha}{\hbar}\,
\Re\left[\,\Psi^*(x,y)\,(\sigma_y \hat{x}-\sigma_x \hat{y})\tau_z\,\Psi(x,y)\,\right]\;.
\end{equation}
In this form it is possible to recover the usual expression for the current in absence of Rashba and superconductivity \cite{Vignale,Beenaker} 
making the Rashba strength $\alpha$ zero and removing $\tau_z$.

\section{Results}
\label{s4}
\begin{SCfigure}
\centering
\resizebox{0.7\textwidth}{!}{
	\includegraphics{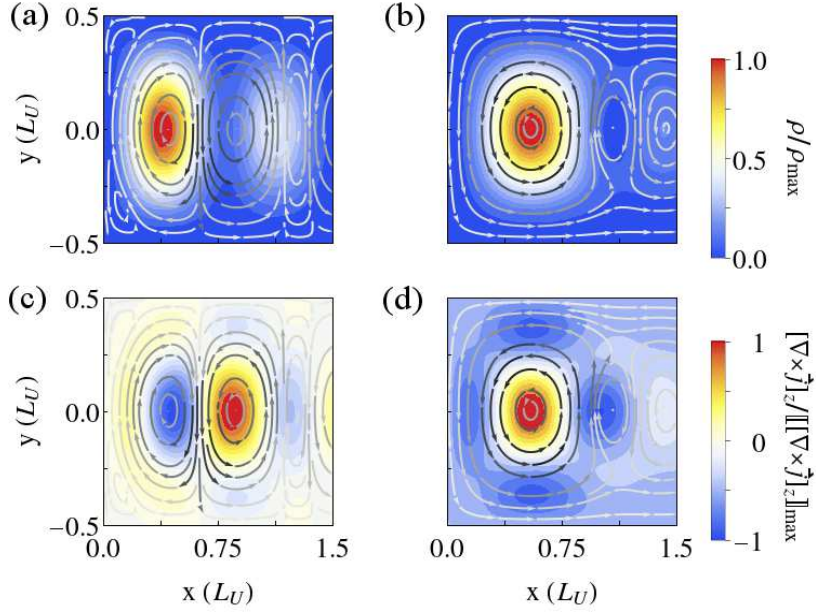}
}
\caption{Current ($\vec{\jmath\,}$) streamlines for longitudinal (a, c) and tilted field
 (b, d). The background in panels (a, b) shows the density, while in (c, d) it shows 
the $z$-component of the rotational $[\nabla\times\vec{\jmath\,}]_z$. In each case the fields
are normalized to their respective maximum values. The parameters are the same of Fig.\ \ref{F1}. The streamlines are colored in a gray scale to show the regions of maximum current in black and the regions of zero current in white.}
\label{F2}
\end{SCfigure}

We can see in Figs.\ \ref{F2}a and \ref{F2}b the quasi-particle current streamlines with the 
Majorana density $\rho(x,y)$ in the background,
for two cases. In the first one (Fig.\ \ref{F2}a) the magnetic field points along the longitudinal direction of the nanowire ($\theta=90^\circ$) while in the second one (Fig.\ \ref{F2}b) the magnetic field has a component perpendicular to the surface of the nanowire ($\theta=75^\circ$). We notice that the current field is modified by the electronic orbital motion in the second case. 

The main effect of the Majorana, in terms of quasi-particle transport, is to create circulating currents. For the longitudinally oriented magnetic field there is no preferred circulation orientation. As shown in Fig.\ \ref{F2}c, the ($z$-component) rotational of the current has a maximum near the Majorana density peak and a minimum between the main and the secondary Majorana peaks.  In the second case, with a perpendicular component of the magnetic field, a circulation orientation is enhanced over the other. We can see a strong current circulation around
the Majorana density peak (see Fig.\ \ref{F2}d) creating a maximum in the rotational near that point. This has similarities to what happens to the current near vortices in p-wave superconductors. On the other hand, the region of high circulating current around the Majorana peak is surrounded by regions of low or no current near the edges of the nanowire. This gives rise to minima in the value of the rotational. 

\section{Conclusions}
We have extended the study of Ref.\ \cite{Osca4} by calculating the local current associated
with the Majorana state. We have shown that the main effect is the emergence of circulating currents.
These currents follow closed trajectories without a preferred direction of circulation, in absence 
of a perpendicular component of the magnetic field. On the other hand, when a perpendicular component is present a
quasi-particle current circulates around the Majorana peak in a vortex-like scenario. These quasi-particle currents
may be difficult to measure experimentally in closed nanowires. However, in future work we will address the study of how these 
quasi-particle currents may change in open Majorana nanowires where they give a contribution to the nanowire conductance.

\section{Acknowledgments}
This work was funded by MINECO-Spain (grant FIS2011-23526), CAIB-Spain (Conselleria d'Educaci\'o,
Cultura i Universitats) and FEDER. We hereby acknowledge the PhD grant provided by the University 
of the Balearic Islands.

\section*{References}
\bibliography{Proceedings}

\providecommand{\newblock}{}
\begin{thebibliography}{10}
\expandafter\ifx\csname url\endcsname\relax
  \def\url#1{{\tt #1}}\fi
\expandafter\ifx\csname urlprefix\endcsname\relax\def\urlprefix{URL }\fi
\providecommand{\eprint}[2][]{\url{#2}}

\bibitem{Alicea}
Alicea J 2012 {\em Rep.\ Prog.\ Phys.\/} {\bf 75} 076501

\bibitem{Beenaker2}
Beenakker C~W~J 2013 {\em Annu.\ Rev.\ Condens.\ Matter Phys.\/} {\bf 4} 113

\bibitem{StanescuREV}
Stanescu T~D and Tewari S 2013 {\em J. Phys.\ Condens.\ Matter\/} {\bf 25}
  233201

\bibitem{Franz}
Franz M 2013 {\em Nature Nanotechnology\/} {\bf 8} 149

\bibitem{Lutchyn2}
Lutchyn R~M, Stanescu T~D and Das~Sarma S 2011 {\em Phys. Rev. Lett.\/} {\bf
  106}(12) 127001

\bibitem{Osca4}
Osca J and Serra L 2015 {\em Phys. Rev. B\/} {\bf 91}(23) 235417

\bibitem{Serra}
Serra L 2013 {\em Phys.\ Rev.\ B\/} {\bf 87} 075440

\bibitem{Molenkamp}
Molenkamp L~W, Schmidt G and Bauer G~E~W 2001 {\em Phys. Rev. B\/} {\bf 64}(12)
  121202

\bibitem{Vignale}
Vignale G and Rasolt M 1988 {\em Phys. Rev. B\/} {\bf 37}(18) 10685--10696

\bibitem{Beenaker}
Beenakker C 1992 {\em Low-Dimensional Electronic Systems\/} ({\em Springer
  Series in Solid-State Sciences\/} vol 111) ed Bauer G, Kuchar F and Heinrich
  H (Springer Berlin Heidelberg) pp 78--82 ISBN 978-3-642-84859-9

\end{thebibliography}

\end{document}